\documentclass[twocolumn,english,conference]{IEEEtran}

\usepackage[cmex10]{amsmath}
\usepackage[margin=0.9in]{geometry}

\usepackage{cite}

\usepackage{latexsym}
\usepackage{color}
\usepackage{graphics}
\usepackage{amssymb}
\usepackage{amsthm}

\usepackage{cite}
\usepackage{array}
\usepackage{graphicx}

\usepackage{url}

\newcommand{\condSum}[3]{\overset{#3}{\underset{\underset{#2}{#1}}{\sum}}}

\newcommand{\vect}[1]{\mathbf{#1}}

\def\Htran{\mbox{\tiny $\mathrm{H}$}}
\def\Ttran{\mbox{\tiny $\mathrm{T}$}}

\theoremstyle{plain}

\newtheorem{theorem}{Theorem}
\newtheorem{corollary}{Corollary}
\newtheorem{lemma}{Lemma}
\newtheorem{definition}{Definition}

\begin{document}

\title{Three Practical Aspects of Massive MIMO: Intermittent User Activity, Pilot Synchronism, and Asymmetric Deployment}

\IEEEoverridecommandlockouts

\author{\IEEEauthorblockN{Emil Bj{\"o}rnson and Erik G.~Larsson}
\IEEEauthorblockA{Department of Electrical Engineering (ISY), Link\"{o}ping University, Link\"{o}ping, Sweden}
\thanks{This research has received funding from the EU FP7 under ICT-619086 (MAMMOET), ELLIIT, and CENIIT.}%
}

\maketitle

\begin{abstract}
This paper considers three aspects of Massive MIMO (multiple-input multiple-output) communication networks that have received little attention in previous works, but are important to understand when designing and implementing this promising wireless technology. First, we analyze how bursty data traffic behaviors affect the system. Using a probabilistic model for intermittent user activity, we show that the spectral efficiency (SE) scales gracefully with reduced user activity. Then, we make an analytic comparison between synchronous and asynchronous pilot signaling, and prove that the choice between these has no impact on the SE. Finally, we provide an analytical and numerical study of the SE achieved with random network deployment.
\end{abstract}

\vspace{-2mm}

\section{Introduction}

\vspace{-1mm}

\label{sec:intro}

Space-division multiple access (SDMA) is a technology for high spectral efficiency (SE) in cellular networks, based on serving multiple user equipments (UEs) in the same time-frequency resource. While the basic concepts behind SDMA are more than 20 years old \cite{Winters1987a,Swales1990a,Richard1996a}, a more thorough understanding of the associated multi-user information theory with multi-antenna base stations (BSs) is a younger science \cite{Caire2003a,Viswanath2003a,Goldsmith2003a}. The ultimately most scalable form of SDMA is Massive MIMO, conceived in \cite{Marzetta2010a}, which exploits the following properties \cite{Marzetta2010a,Huh2012a,Larsson2014a}: 1) Having many more antennas than UEs give high spatial separability and near optimality of linear processing; 2) Channel state information (CSI) is only needed at the BS and can be obtained by uplink pilot signaling in time-division duplex (TDD) mode; 3) Channel hardening suppresses frequency/small scale fading, which makes the power allocation easier and eliminates the need for downlink pilots. The analytical study \cite{Bjornson2016a} has recently shown that Massive MIMO can provide 50-fold gains in SE over IMT-Advanced, without any BS cooperation, thus making it a strong candidate technology for next generation wireless technologies \cite{Boccardi2014a}.

As the basic communication theoretic performance analysis and asymptotic limits of Massive MIMO are becoming mature, it is particularly important to consider practical implementation aspects of the technology. The first real-time Massive MIMO testbed was described in \cite{Vieira2014a} and it shows that the Massive MIMO technology really works.

In this paper, we consider three aspects of Massive MIMO that are important in practical implementations:

\begin{itemize}
\item \textbf{Intermittent user activity:} Most prior studies on Massive MIMO consider full-buffer transmission, where each UE transmits/receives data continuously. In contrast, practical higher layers can create burstiness (even under data streaming), thus there is not always data to transmit. We take a first step towards a more realistic traffic modeling by deriving a new SE expression, for a particular model of intermittent user activity, and analyze the performance implications.

\item \textbf{Synchronous versus asynchronous pilot signaling:} 

Since the seminal paper \cite{Marzetta2010a} by Marzetta, synchronous pilot transmission across all cells has been the common practice, in an attempt to reduce the inter-cell interference during pilot transmission. There are several reasons to question this basic assumption. Firstly, time-synchronous transmission is hard to obtain over a large coverage area. Secondly, coordinated per-user pilot allocation is cumbersome between cells since the set of active UEs changes by scheduling and intermittent user activity, while fixed pilot reuse patterns with random pilot allocation within each pilot cluster is a more practical solution \cite{Bjornson2016a,Atzeni2015a}. Thirdly, \cite{Ngo2013a} and \cite{Fernandes2013a} showed that having unsynchronized pilot transmission will not cause any fundamental differences. We compare synchronous and asynchronous pilot signaling, and obtain exciting new analytical results when using maximum ratio (MR) combining.

\item \textbf{Performance analysis in asymmetric deployments:} The resource allocation papers \cite{Huh2012a} and \cite{Bjornson2016a} have analyzed the anticipated performance in symmetrically deployed Massive MIMO networks. In contrast, we analyze the SE in randomly deployed uncoordinated networks where the BSs are distributed according to a Poisson point process (PPP), which models the randomness of practical network deployments \cite{Andrews2011a}. Closed-form SE expressions and limits are derived.

\end{itemize}

\begin{figure*}[t!] 
\begin{equation} \label{eq:SINR-value} \tag{3}
\mathrm{SINR}_{0k}  \!=\!  \frac{ p_{0k} | \mathbb{E}\{  \vect{h}_{00k}^{\Htran} \vect{g}_{0k} \} |^2 }{ \!
 \sum\limits_{i=1, i \neq k}^{K} p_{0i}  \mathbb{E} \{  | a_{0i} \vect{h}_{00i}^{\Htran} \vect{g}_{0k} |^2  \} + p_{0k}  \mathbb{V} \{ \vect{h}_{00k}^{\Htran}  \vect{g}_{0k} \} \! + \!\! \sum\limits_{j \in \Psi} \sum\limits_{i=1}^{K} p_{ji}  \mathbb{E} \{ | a_{ji} \vect{h}_{0ji}^{\Htran} \vect{g}_{0k}  |^2  \}   + \sigma^2  \mathbb{E} \{ \| \vect{g}_{0k} \|^2\}  }
\end{equation} \vskip-1mm
\hrulefill \vspace{-5mm}
\end{figure*}

\section{System Model}

\label{sec:system}

We consider the uplink of a cellular network where each BS is equipped with $M$ receive antennas and serves $K$ single-antenna UEs. Since the uplink performance of a cell depends on the receive processing performed at the local BS, but is independent of the processing at other BSs, we can without loss of generality concentrate on $\mathrm{BS}_0$ with some arbitrary cell index $0$. This cell is surrounded by interfering cells with indices in some arbitrary set $\Psi$.

The uplink time-frequency resources are divided into blocks of $T_c$ seconds and $W_c$ Hz, which leaves room for $S = T_c W_c$ transmission symbols for pilot signaling and payload data. The channel response $\vect{h}_{ljk} \in \mathbb{C}^M$ between $\mathrm{BS}_l$ and UE $k$ in cell $j$ is modeled as block fading, thus $\vect{h}_{ljk}$ is constant within a block and takes independent random realizations across blocks. More precisely, we consider circularly symmetric complex Gaussian realizations $\vect{h}_{ljk} \sim \mathcal{CN}(\vect{0}, \beta_{ljk}  \vect{I}_M)$, where $\beta_{ljk} >0 $ is the average channel attenuation between UE $k$ in cell $j$ and any of the antennas at $\mathrm{BS}_l$. This is known as Rayleigh block fading.

Since continuous data services can give rise to intermittent data bursts, we assume that each UE has a probability $A$ of being active in a given coherence block, where $0 \leq A \leq 1$.\footnote{With this model, the number of active UEs in a cell in a given block coherence has a binomial distribution: $\mathcal{B}(K,A)$.} This is a key difference from prior works were only the special case $A=1$ has been considered. The intermittent user activity of  UE $i$ in cell $j$ is represented by a binary random variables $a_{ji} \in \{0,1\}$, where $a_{ji}=1$ if the UE is active in the current block and $a_{ji}=0$ stands for inactivity. These events occur with probability $A$ and $1-A$, respectively, and are independent between blocks.

The received signal $\vect{y}_0 \in \mathbb{C}^{M}$ at $\mathrm{BS}_0$ in an arbitrary coherence block is then modeled as
\begin{equation} \label{eq:system-model-received}
\vect{y}_0 = \sum_{i=1}^{K}  \vect{h}_{00i} a_{0i} s_{0i} + \sum_{j \in \Psi}
\sum_{i=1}^{K} \vect{h}_{0ji} a_{ji} s_{ji} + \vect{n}_{0}
\end{equation}
where the symbol $s_{ji} \in \mathbb{C}$ transmitted by UE $i$ in cell $j$ has the average energy $p_{ji} =\mathbb{E}\{ | s_{ji} |^2 \} $ and $\vect{n}_{0} \sim \mathcal{CN}(\vect{0}, \sigma^2 \vect{I}_M)$ denotes  temporally white noise.

\section{SE under Intermittent User Activity}
\label{eq:performance-general-case}

The considered $\mathrm{BS}_0$ applies a linear combining vector $\vect{g}_{0k} \in \mathbb{C}^{M}$, as $ \vect{y}_0^{\Htran} \vect{g}_{0k}$, to detect the signal sent by its $k$th UE, when this UE is active (i.e., $a_{0k}=1$). The next lemma gives an achievable SE under intermittent user activity.

\begin{lemma} \label{lemma:SE}
If $\mathrm{BS}_0$ knows when its $k$th UE is active, a lower bound on its ergodic capacity is
\begin{equation}
\mathrm{SE}_{0k} = A \left( 1-\frac{B}{S} \right)  \log_2 \left(1+ \mathrm{SINR}_{0k} \right) \,\,\,\, \text{[bit/s/Hz]}
\end{equation}
where the effective signal-to-interference-and-noise ratio (SINR) is given in \eqref{eq:SINR-value} at the top of the page and the expectations $\mathbb{E} \{ \cdot \}$ and variance $\mathbb{V} \{ \cdot \}$ are taken with respect to channel realizations, inter-cell interference, and user activation variables (conditioned on $a_{0k}=1$).
\end{lemma} \vskip-2mm
\begin{IEEEproof}
This is obtained by lower bounding the mutual information in the same way as in \cite[Lemma 2]{Bjornson2016a}, while taking the user activity probabilities into account.
\end{IEEEproof}

\setcounter{equation}{3}

The ergodicity of the spectral efficiency $\mathrm{SE}_{0k}$ is with respect to the random channel realizations and user activities. The SE depends on the combining vector $\vect{g}_{0k}$, which ideally should be matched to the current channel realizations; for example, to maximize the signal gain $| \mathbb{E}\{  \vect{h}_{00k}^{\Htran} \vect{g}_{0k} \} |^2$. Channel estimation is thus very important.

\begin{figure*}
\begin{equation} \label{eq:SINR-value-closed-form} \tag{10}
\mathrm{SINR}_{0k}  \!=\!  \frac{ M \beta_{00k}^2 }{ \! \left( \! \beta_{00k} + \sum\limits_{j \in \Psi } \sum\limits_{i=1}^{K}  \! \frac{ A p_{ji}   \beta_{0ji} }{p_{0k} B}  + \frac{ \sigma^2}{p_{0k} B} \!\right) \!\!\! \Bigg(\! \beta_{00k} +
 \condSum{i=1}{i \neq k}{K} \frac{A p_{0i}   \beta_{00i}}{p_{0k}}  +     \sum\limits_{j \in \Psi} \sum\limits_{i=1}^{K} \frac{A p_{ji}   \beta_{0ji}}{p_{0k}}  + \frac{\sigma^2}{p_{0k}} \!\Bigg) \!+ \!\sum\limits_{j \in \Psi} \sum\limits_{i=1}^{K} \frac{ A p_{ji}^2  \beta_{0ji}^2 \left(  M + 1-A    \right) }{p_{0k}^2 B}    }
\end{equation} \vskip-2mm
\hrulefill \vspace{-4mm}
\end{figure*}

\vskip-2mm

\subsection{Channel Estimation and Maximum Ratio Combining}

\vskip-1mm

As described in Section \ref{sec:intro}, we will compare synchronous and asynchronous pilot transmission to gain new insights into the performance difference between these approaches. Each active UE in cell $0$ sends a pilot sequence of length $B$ over $B$ instances of the system model in \eqref{eq:system-model-received}, where $B \geq K$ to allow for orthogonal sequences when all UEs are active. Let $\vect{s}_{ji} =  [s_{ji}^{1} \, \ldots \, s_{ji}^{B} ]^{\Ttran} \in \mathbb{C}^{B}$ denote the combined signal sent by UE $i$ in cell $j$ during the pilot transmission in cell $0$. For UE $k$ in cell $0$, its signal $s_{0k}$ is a pilot sequence $\vect{s}_{0k} = \sqrt{p_{0k}}\vect{v}_{k}$, where $\vect{v}_{k}$ is selected from a pilot book \vskip-2mm
\begin{equation}
\mathcal{V} = \{ \vect{v}_1, \ldots, \vect{v}_B \}
\end{equation} 
\noindent where $\| \vect{v}_b \|^2=B$, for $b=1,\ldots,B$, and the sequences are orthogonal in the sense that $\vect{v}_{b_1}^{\Htran} \vect{v}_{b_2} = 0$ for $b_1 \neq b_2$.

The combined received signal from the pilot transmission at $\mathrm{BS}_0$  is denoted as $\vect{Y}_0 \in \mathbb{C}^{M \times B}$ and given by    \vskip-2mm
\begin{equation} \label{eq:Y0}
\begin{split}
\vect{Y}_0 \!=\! \sum_{i=1}^{K} \sqrt{p_{0i}} a_{0i} \vect{h}_{00i} \vect{v}_{i}^{\Ttran}  + \sum_{j \in \Psi}
\sum_{i=1}^{K} a_{ji} \vect{h}_{0ji}  \vect{s}_{ji}^{\Ttran} + \vect{N}_{0}
\end{split}
\end{equation} \vskip-1mm
\noindent where $\vect{N}_{0} \in \mathbb{C}^{M \times B}$ has independent $\mathcal{CN}( 0, \sigma^2 )$-entries. Note that an inactive UE does not send pilots nor data.

The interfering uplink signals $\vect{s}_{ji} \! \in \! \mathbb{C}^{B}$, $j \! \neq \!0$, are different for synchronous and asynchronous pilot transmission.

\begin{definition}[Synchronous pilots]
All cells send pilots simultaneously. In each coherence block, $\mathrm{BS}_j$ selects $K$ different pilots from $\mathcal{V} $ uniformly at random, such that $\vect{s}_{ji} \!=\! \sum_{b=1}^{B}  \sqrt{p_{ji}}  \chi_{ji}^{b} \vect{v}_{b}$ where $\chi_{ji}^{1},\ldots,\chi_{ji}^{B} \in \{0,1\}$ are binary random variables. These variables satisfy $\sum_{b=1}^{B} \chi_{ji}^{b} = 1$ and $\chi_{ji}^{b}=1$ occurs with probability $1/B$. Since at most one UE per cell uses each pilot, we have $\sum_{i=1}^{K} \chi_{ji}^{b} \leq 1$.
\end{definition}

\begin{definition}[Asynchronous pilots]
The interfering UEs transmit any random signals $\vect{s}_{ji} = [s_{ji}^{1} \, \ldots \, s_{ji}^{B} ]^{\Ttran}$ (for $i=1,\ldots, K$ and $j \in \Psi$), where $s_{ji}^{b}$ is an independent random variable with the average energy $\mathbb{E}\{ | s_{ji}^{b} |^2 \} = p_{ji}$.
\end{definition}
When UE $k$ in cell $0$ is active (i.e., $a_{0k}=1$), $\mathrm{BS}_0$ can correlate $\vect{Y}_0$ in \eqref{eq:Y0} with the $k$th pilot sequence  to obtain
\vspace{-2mm}
\begin{align}  \notag
\vect{Y}_0 \vect{v}_{k}^* &=  \! \sqrt{p_{0k}} \vect{h}_{00k} B + \! \sum_{j \in \Psi }
\sum_{i=1}^{K}  \! a_{ji} \vect{h}_{0ji}  \vect{s}_{ji}^{\Ttran}\vect{v}_{k}^* + \vect{N}_{0} \vect{v}_{k}^* \\
& \quad \quad \textrm{for} \quad k \in \{1,\ldots, K\},\label{eq:sufficent-statistics}
\end{align}
where $\vect{N}_{0} \vect{v}_{k}^* \sim \mathcal{CN}(\vect{0},B \sigma^2 \vect{I}_M )$.
Clearly, $\vect{Y}_0 \vect{v}_{k}^*$ is a sufficient statistic for estimating $\vect{h}_{00k}$. Since Bayesian estimators require known statistics of the interfering signals, which can be hard to obtain under intermittent user activity, we consider the least-square (LS) estimator \vskip-2mm
\begin{equation} \label{eq:LS-estimator}
  \hat{\vect{h}}_{00k} = \frac{a_{0k}}{\sqrt{p_{0k}} B} \vect{Y}_0 \vect{v}_{k}^*.
\end{equation} \vskip-2mm
\noindent Note that if UE $k$ is inactive (i.e., $a_{0k} = 0$) then $\hat{\vect{h}}_{00k}  = \vect{0}$ since $\vect{Y}_0 \vect{v}_{k}^*$ only contains interference and noise.

We assume that $\mathrm{BS}_0$ utilizes the LS channel estimate in \eqref{eq:LS-estimator} for MR combining where  $\vect{g}_{0k} = \hat{\vect{h}}_{00k}$.
This receive combining scheme attempts to maximize the desired signal gain $| \mathbb{E}\{  \vect{h}_{00k}^{\Htran} \vect{g}_{0k} \} |^2$ in Lemma \ref{lemma:SE}. With MR we can now obtain closed-form expressions for the achievable SEs in Lemma \ref{lemma:SE}, for both synchronous and asynchronous pilots.

\begin{theorem} \label{theorem:SINR-MRC}
For both synchronous and asynchronous pilot transmission, the lower bound on the ergodic capacity of UE $k$ in cell $0$
in Lemma \ref{lemma:SE} becomes\begin{equation} \label{eq:ergodic-capacity}
A \Big( 1 - \frac{B}{S} \Big) \log_2 \! \left( 1 + \mathrm{SINR}_{0k}  \right) \quad \textrm{[bit/s/Hz]} 
 \end{equation}
for MR combining, where the effective SINR, $\mathrm{SINR}_{0k}$, is given in \eqref{eq:SINR-value-closed-form} at the top of the page.
\end{theorem}
\vspace{-1mm}
\begin{IEEEproof}
The proof is outlined in the appendix.
\end{IEEEproof}

\setcounter{equation}{10}

Interestingly, this theorem shows that the choice between synchronous and asynchronous pilot transmission has no impact on the SE in Lemma \ref{lemma:SE}, because the average interference during pilot transmission is suppressed by a factor $1/B$ in both cases. In the synchronous case this is attributed to the $1/B$ probability of using the same pilot, while the incoherent inter-cell signaling in the asynchronous case gives the same interference suppression. Coding redundancy  cannot reduce pilot contamination, only increase $B$. The fact that also interfering data transmissions cause pilot contamination was previously demonstrated in \cite{Ngo2013a,Fernandes2013a}, thus the novel contribution of Theorem \ref{theorem:SINR-MRC} is that the impact on the classic SE lower bound is exactly equal. Our numerical results hold for both cases.

\begin{figure*}
\begin{equation} \label{eq:SINR-value-stochastic} \tag{13}
\underline{\mathrm{SINR}} \!=\!  \frac{ M }{ \frac{  K^2 }{ B} \left( \frac{4 A^2}{(\alpha-2)^2} + \frac{A^2}{\alpha-1} \right)  + \frac{2K A}{\alpha-2} \left(   1 +
 \frac{ 1 }{ B} +
  \frac{ A (K-1)}{ B}  +  \frac{2 \sigma^2}{ \rho B}   \right) \!+\! \left(  1   + \frac{ \sigma^2}{ \rho B} \right) \!\!\left( 1 + (K-1)A   + \frac{\sigma^2}{\rho} \right) 
+  \frac{ A    \left(  M + 1-A    \right) }{ B} \frac{K}{\alpha-1}   }
\end{equation} \vskip-2mm
\hrulefill \vspace{-4mm}
\end{figure*}

\vspace{-1mm}

\section{Average Performance in Random Networks}

\begin{figure} \vskip-2mm
\begin{center}
\includegraphics[width=\columnwidth]{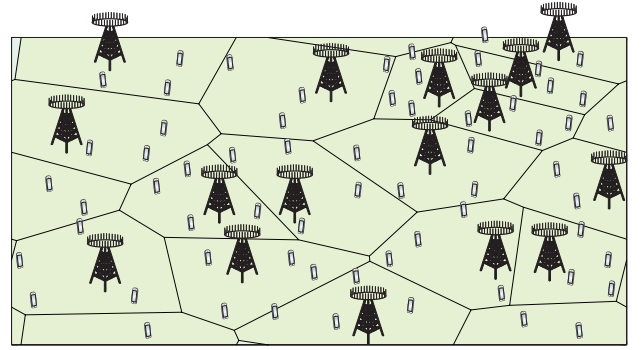}
\end{center} \vskip-5mm
\caption{Example of BS positions from a homogeneous PPP $\Phi$ and UEs uniformly distributed in the corresponding Poisson-Voronoi cells.} \label{figureVoronoi} \vskip-3mm
\end{figure}

The performance analysis in Section \ref{eq:performance-general-case} holds for the uplink of any Massive MIMO network deployment. To obtain more quantitative results, we now consider a random BS deployment based on stochastic geometry. More precisely, we assume that the BSs are distributed in $\mathbb{R}^2$ according to a homogeneous PPP $\Phi$ with intensity $\lambda$ BSs per $\textrm{km}^2$. This means that in any sub-area of size $C$ $\textrm{km}^2$ there are $\mathrm{Po}(\lambda C)$ uniformly and independently distributed BSs, where $\mathrm{Po}(\cdot)$ is a Poisson distribution. Each UE connects to its closest BS, thus the coverage area of a BS is its Poisson-Voronoi cell; see Fig.~\ref{figureVoronoi}. The key features of this model are the asymmetry and that the BSs are matched to a spatially heterogeneous user load, where small cells have higher UE density (measured per $\textrm{km}^2$) than larger cells.

The goal of this section is to characterize the SE per cell in the network, which can be done by studying the statistical properties of any UE, because each UE is statistically ``typical'' due to the translation invariance of homogeneous PPPs \cite{Weber2010a}. Using the same notation as in previous sections, we focus on a \emph{typical UE} with index $k$ that is connected to the typical BS $\mathrm{BS}_0 \in \Phi$. The following result is obtained from basic PPP properties (see \cite{Weber2010a}):

\begin{lemma} \label{lemma:distance-distribution}
The distance $d_{00k}$ between the typical UE and $\mathrm{BS}_0$ is distributed as $d_{00k} \sim \mathrm{Rayleigh} \big( \frac{1}{\sqrt{2\pi \lambda}} \big)$. The BSs of other cells form a homogenous PPP $\Psi = \Phi \setminus \{ \mathrm{BS}_0 \}$ in $\mathbb{R}^2 \setminus \{ \vect{x} \in \mathbb{R}^2 : \| \vect{x} \| < d_{00k} \}$.
\end{lemma}

The propagation distance between $\mathrm{BS}_j$ and UE $k$ in cell $j$ is denoted by $d_{ljk}$ (in km). The corresponding channel attenuation is assumed to be distant-dependent as $\beta_{ljk} = \omega^{-1} d_{ljk}^{-\alpha}$, where $\alpha> 2$ is the pathloss exponent and $\omega$ determines the average channel attenuation at a reference distance of 1 km. Since this model causes large channel quality variations within the cells, uplink power control is necessary to avoid near-far blockage where weak signals from distant UEs drown in stronger signals from nearby UEs. We consider statistical channel inversion where UE $i$ in cell $j$ uses the average symbol energy
\begin{equation}\label{eq:power-control}
p_{ji} = \rho/\beta_{jji}
\end{equation}
where $\rho \geq 0$ is a design variable and $\beta_{jji}$ is the channel attenuation from the serving BS. This power control policy provides the same average channel gain $\mathbb{E}\left\{{p_{ji} \|\vect{h}_{jji}\|^2} \right\} = M {\rho}$ to all UEs in the network, without having to feedback any instantaneous channel realizations \cite{Bjornson2016a}. The maximum output power of the UEs limits the choice of $\rho$ in practice.

Under these assumptions, we compute a tight lower bound on the average of the SE expression in Theorem~\ref{theorem:SINR-MRC}.

\begin{theorem} \label{theorem:average-SE}
A lower bound on the average uplink capacity [bit/symbol/user] is
\begin{equation} 
\underline {\mathrm{SE}} = A \Big( 1 - \frac{B}{S} \Big) \log_2  \left( 1 + \underline{\mathrm{SINR}} \right)\label{eq:average-SE}
\end{equation}
where $\underline{\mathrm{SINR}}$ is given in \eqref{eq:SINR-value-stochastic} at the top of the page.
\end{theorem}
\begin{IEEEproof}
The proof is outlined in the appendix.
\end{IEEEproof}

\setcounter{equation}{13}

The closed-form average SE in Theorem \ref{theorem:average-SE} depends on a number of parameters: $A$, $B$, $M$, $K$, $\alpha$, and $\rho/\sigma^2$. Interestingly, the power control policy makes the SE independent of the BS density $\lambda$ and of $\omega$, but the average energy per symbol depends on $\lambda$ as $\mathbb{E}\{p_{ji} \}= \mathbb{E}\{ \frac{\rho}{\beta_{jji}} \} = \rho \omega \Gamma(\alpha/2-1)/ (\pi \lambda)^{\alpha/2}$, where $\Gamma(\cdot)$ is the Gamma function. Most of these parameters are decided by the deployment scenario, but the number of pilot sequences, $B$, can be optimized.

\begin{corollary} \label{cor:asymptotic-limit}
The average uplink SE in Theorem \ref{theorem:average-SE} is a  concave function of $B$, having its unique global maximum somewhere in the range $K \leq B \leq S$.
\end{corollary}
\begin{IEEEproof}
The average SE has the structure $A (1-B/S) \log_2(1+1/(C_1+C_2/B))$, for some constants $C_1,C_2 \geq 0$, which is a concave function of $B$.
\end{IEEEproof}

This corollary shows that there is a potential benefit of using pilot sequences with a length $B$ that is longer than $K$, to suppress inter-cell interference under both synchronous and asynchronous pilot transmission. The asymptotic behavior is given by the next corollary.

\begin{corollary} \label{cor:asymptotic-B}
The average uplink SE in Theorem \ref{theorem:average-SE} behaves as
\begin{equation}  \label{eq:SE-limit}
\underline {\mathrm{SE}} \rightarrow A \Big( 1 - \frac{B}{S} \Big) \log_2  \left( 1 + \frac{B(\alpha-1)}{AK} \right)
\end{equation}
when $M \rightarrow \infty$. This asymptotic limit is maximized by
\begin{equation}
B = 
\frac{ \frac{AK}{\alpha-1} +S}{W \left( e \left(1+ \frac{S(\alpha-1)}{AK}  \right) \right)} 
- \frac{AK}{\alpha-1} 
\end{equation}
where the number $e$ is the base of natural logarithm and $W(\cdot)$ is the Lambert $W$ function.\footnote{The Lambert $W$ function is defined implicitly via $x=W(x) e^{W(x)}$.}
\end{corollary}
\begin{IEEEproof}
This follows from taking the limit $M \rightarrow \infty$ in \eqref{eq:SINR-value-stochastic} and then finding the point where the first derivative of the concave SE function in \eqref{eq:SE-limit} is zero.
\end{IEEEproof}

\subsection{Numerical Examples}

We illustrate the anticipated SE in random deployments by a numerical study with $K=30$ UEs per cell and either $M=100$ or $M=500$ BS antennas. The asymptotic limit, $M \rightarrow \infty$, from Corollary \ref{cor:asymptotic-limit} is also considered. The pathloss exponent is selected as $\alpha = 3.76$, which is a common number for urban 3GPP simulations. We set the SNR, $\rho/\sigma^2$, to 5 dB, which is enough to reach the interference-limited regime in Massive MIMO networks \cite{Bjornson2016a}. Each coherence block contains $S=400$ symbols (e.g., achieved by $T_c = 2$ ms and $W_c = 200$ kHz).

\begin{figure}
\begin{center}
\includegraphics[width=\columnwidth]{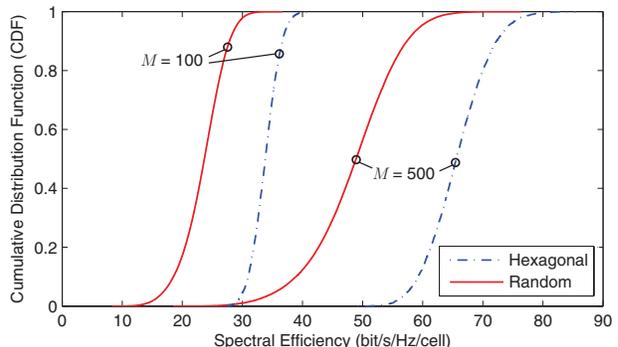}
\end{center} \vskip-6mm
\caption{Empirical cumulative distribution function of the SE per cell, for either hexagonal and random network deployment.} \label{figure_cdf} \vskip-6mm
\end{figure}

We first compare random network deployment with symmetric hexagonal cell deployment in terms of the SE per cell.
 Fig.~\ref{figure_cdf} shows the empirical cumulative distribution functions (CDFs) with respect to user locations (and BS locations for random deployment). The curves were generated by Monte-Carlo simulations based on Theorem~\ref{theorem:SINR-MRC} with $M \in \{100, \, 500\}$, $A=1$ and $B = 60$, while we note that the average SE with random deployment is lower-bounded by Theorem \ref{theorem:average-SE}. Despite the considered power-control, which provides all UEs with equal SNR, there are large SE variations since the inter-cell interference depends the location of the other-cell UEs. The SE variations are smaller with fewer antennas since the interference is more dominant. Hexagonal cells provide higher average SE than random deployment ($+44$\% at $M=100$ and $+36$\% at $M=500$) and gives CDFs with shorter tails, since there are often more neighboring cells in random deployment.
Hence, the BS deployment has a major impact on the SE.

\begin{figure}
\begin{center}
\includegraphics[width=\columnwidth]{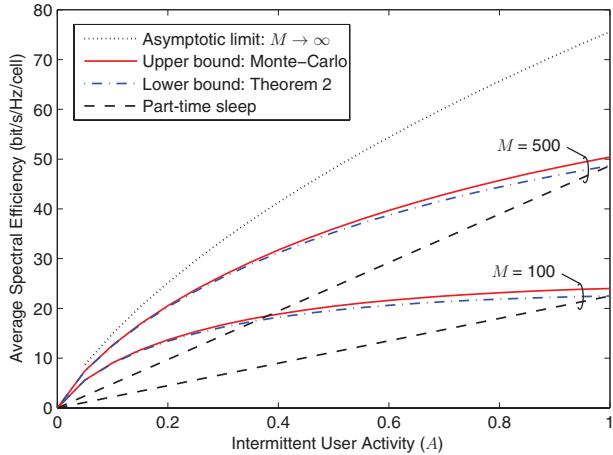}
\end{center} \vskip-4mm
\caption{Average SE per cell, for optimized pilot lengths, as a function of the intermittent user activity (i.e., the percentage of coherence blocks were each UE has data to transmit). There are $K=30$ UEs per cell.} \label{figure_activity} \vskip-6mm
\end{figure}

Fig.~\ref{figure_activity} shows the average SE per cell, over different random BS and UE positions, as a function of the  user activity $A$. The pilot length $B$ is optimized numerically to give maximal SE. The simulation is based on the average of the SE in Theorem \ref{theorem:SINR-MRC}; Fig.~\ref{figure_activity} shows both the closed-form lower bound in Theorem \ref{theorem:average-SE} and Monte-Carlo simulations that lead to an upper bound due to a finite number of interfering cells. The small gap between the curves show the tightness of the closed-form expression in Theorem \ref{theorem:average-SE}.

The average SE in Fig.~\ref{figure_activity} decreases with the user activity, since each UE is only active in a fraction $A$ of the blocks. However, the performance loss is not proportional to the user activity, but often much smaller. This can be seen by comparing the lower bound with the line ``part-time sleep'' that shows the case where only a fraction $A$ of all coherence blocks are used but all $K$ UEs are active in these blocks. The reason is that the network is interference-limited so the loss in per-cell SE from only having, on average, $AK \leq K$ active UEs in the cell is partially compensated by the lower interference from having, on average, $AK \leq K$ interfering UEs in the other cell. In other words, intermittent user activity seems to have a limited affect on the performance; for example, with $M=100$ antennas a $30 \%$ user activity only gives a $32 \%$ performance loss, and not $70 \%$ as one might expect.

Adding more BS antennas improves the average SE in Fig.~\ref{figure_activity}. The gap and relative loss as compared to the asymptotic limit ($M \rightarrow \infty$) is smaller for low user activity probabilities, which is encouraging for practical implementation---fewer antennas seem to be needed to reach a certain percentage of the asymptotic performance. The system achieves $30$\%--$64$\% of the limit with $M\!=\!100$, depending on $A$, and $64$\%--$85$\% of the limit with $M\!=\!500$.

\begin{figure}
\begin{center}
\includegraphics[width=\columnwidth]{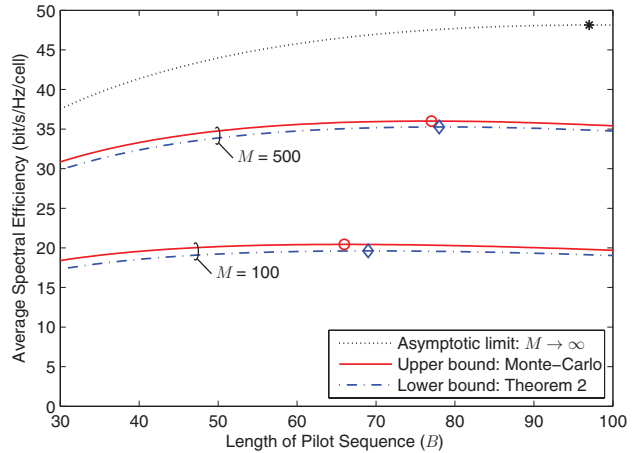}
\end{center} \vskip-4mm
\caption{Average SE per cell, for $A=0.5$, as a function of the pilot sequence length. There are $K=30$ UEs per cell. The maximum value of each curve is marked.}\label{figure_pilotsequence} \vskip-2mm
\end{figure}

The impact of the pilot sequence length, $B$, is illustrated in Fig.~\ref{figure_pilotsequence}. The average SE per cell is shown as a function of $B$, for $A=0.5$ and different number of antennas. The optimal $B$ is marked on each curve and, interestingly, it lies in the range  $60$--$100$ even if there is only $K=30$ UEs per cell. Hence, it is preferable to have 2--3 times more orthogonal pilots than the total number of UEs per cell, even if each cell on average only needs $AK = 15$ of them. The performance variations with respect to $B$ are quite small, thus any selection of $B$ in this range $60$--$100$ works well. Fig.~\ref{figure_pilotsequence} shows that the optimal $B$ increases with $M$. 
The asymptotic limit ($M \rightarrow \infty$) is maximized at $B^* = 97$, which matches the value provided by Corollary~\ref{cor:asymptotic-B}.

\section{Conclusions}
\label{sec:conclusion}

We have taken a look at Massive MIMO from three important practical perspectives, which have been greatly overlooked in previous analytical works.

First, we modeled the natural data traffic variations by a probabilistic intermittent user activity model, where each UE is on average only active in a fraction of the blocks. The analysis shows that reducing the user activity shrinks both the pre-log factor and the average interference. Since the latter is a positive effect, the loss in SE is smaller than one might expect. In fact, the simulations show that one needs fewer antennas to come close to the asymptotic limits under intermittent user activity.

Next, we made an analytic comparison of synchronous and asynchronous pilot transmission and proved that it has no impact on the uplink performance. This confirms previous observations and is good news since it reduces the need for inter-cell pilot coordination. The result was proved for MR combining with random pilot allocation, but similar results are expected with zero-forcing processing and in the downlink. Different results might be obtained if well-designed pilot reuse patterns are considered. Nevertheless, our results show that it is the existence of inter-cell interference during pilot transmission that is the main pilot contamination issue, while the structure of this interference is less important. The pilot contamination effect can thus be analyzed and evaluated in Massive MIMO implementations by sending any interfering signal.

Finally, we analyzed the performance in randomly deployed networks, where the BS locations are given by a PPP. A tight lower bound on the average SE was computed in closed form and the limit of many BS antennas was characterized analytically. Simulations were provided to show how random deployment, intermittent user activity, pilot lengths, and number of antennas affect the system. Random deployment provides substantially lower performance than hexagonal cells, thus showing the importance of coordinating the deployment of Massive MIMO BSs.

\appendices

\section{Outline of Proofs}

\textbf{Proof of Theorem \ref{theorem:SINR-MRC}.} The expectations in the SINR expression \eqref{eq:SINR-value} need to be computed for $\vect{g}_{0k} = \hat{\vect{h}}_{00k}$ in the two pilot transmission cases. We begin by computing
\begin{equation} \label{eq:compute-norm-hhat}
\begin{split}
& \mathbb{E} \{ \| \hat{\vect{h}}_{00k} \|^2\} = M \beta_{00k} \\  &+ \sum_{j \in \Psi } \sum_{i=1}^{K}  \! \frac{ A M  \beta_{0ji} }{p_{0k} B^2} \mathbb{E} \{ |  \vect{s}_{ji}^{\Ttran}\vect{v}_{k}^* |^2  \} + \frac{M \sigma^2}{p_{0k} B}
\end{split}
\end{equation}
by utilizing  \eqref{eq:LS-estimator}, the independence of the channels, and the fact that we conditioned on  $a_{0k}=1$. For the same reasons, the desired signal gain becomes
\begin{equation}
|\mathbb{E}\{ \vect{h}_{00k}^{\Htran}  \hat{\vect{h}}_{00k} \}|^2 = | \mathbb{E}\{ \vect{h}_{00k}^{\Htran}  \vect{h}_{00k} \} |^2= M^2 \beta_{00k}^2,
\end{equation}
from which we also obtain the variance $\mathbb{V}\{ \vect{h}_{00k}^{\Htran}  \hat{\vect{h}}_{00k} \} = \mathbb{E}\{ |\vect{h}_{00k}^{\Htran}  \hat{\vect{h}}_{00k} |^2\} - |\mathbb{E}\{ \vect{h}_{00k}^{\Htran}  \hat{\vect{h}}_{00k} \}|^2$ by noting that
\begin{equation}
\mathbb{E}\{ | \vect{h}_{00k}^{\Htran}  \hat{\vect{h}}_{00k} |^2 \} = \beta_{00k} \mathbb{E} \{ \| \hat{\vect{h}}_{00k} \|^2\}  + M^2 \beta_{00k}^2
\end{equation}
after some simple algebra. Next, the intra-cell interference terms, with $i \neq k$, becomes
\begin{equation}
 \mathbb{E} \{  | a_{0i} \vect{h}_{00i}^{\Htran} \hat{\vect{h}}_{00k}  |^2  \} = A \beta_{00i} \mathbb{E} \{ \| \hat{\vect{h}}_{00k} \|^2\} 
 \end{equation}
since the intra-cell pilot sequences are orthogonal.
 
The inter-cell interference terms are computed as
\begin{equation}
\begin{split}
& \mathbb{E} \{ | a_{ji} \vect{h}_{0ji}^{\Htran} \hat{\vect{h}}_{00k}  |^2  \} = A \beta_{0ji} \mathbb{E} \{ \| \hat{\vect{h}}_{00k} \|^2\} \\ &+ \frac{ A M \beta_{0ji}^2  (M+1-A)  }{p_{0k} B^2} \mathbb{E} \{ |  \vect{s}_{ji}^{\Ttran}\vect{v}_{k}^* |^2  \} 
\end{split}
\end{equation}
by keeping in mind that $\vect{h}_{0ji}$ appears in the expression for $\hat{\vect{h}}_{00k}$. The terms $\mathbb{E} \{ |  \vect{s}_{ji}^{\Ttran}\vect{v}_{k}^* |^2  \} $ depend on whether synchronous and asynchronous pilots are considered. In the synchronous case we have 
\begin{equation} \label{eq:pilot-case-sync}
\mathbb{E} \{ |  \vect{s}_{ji}^{\Ttran}\vect{v}_{k}^* |^2  \} = \frac{p_{ji}}{B} |  \vect{v}_{k}^{\Ttran}\vect{v}_{k}^* |^2 = p_{ji} B,
\end{equation}
since an active UE $i$ in cell $j$ uses the $k$th pilot sequence with probability $1/B$. In the asynchronous case we notice that 
\begin{equation} \label{eq:pilot-case-async}
\mathbb{E} \{ |  \vect{s}_{ji}^{\Ttran}\vect{v}_{k}^* |^2  \} = \vect{v}_{k}^{\Ttran} \mathbb{E} \{   \vect{s}_{ji}^{\Ttran} \vect{s}_{ji}^{*} \}   \vect{v}_{k}^* =  p_{ji} \vect{v}_{k}^{\Ttran} \vect{I}_B  \vect{v}_{k}^* = p_{ji} B,
\end{equation}
since $ \vect{s}_{ji}$ contains independent symbols with average energy $p_{ji}$.
We notice from \eqref{eq:pilot-case-sync} and  \eqref{eq:pilot-case-async} that $\mathbb{E} \{ |  \vect{s}_{ji}^{\Ttran}\vect{v}_{k}^* |^2  \} = p_{ji} B$ in both cases, thus the SINR expression is the same. Finally, we obtain \eqref{eq:SINR-value-closed-form} by plugging \eqref{eq:compute-norm-hhat}--\eqref{eq:pilot-case-async} into \eqref{eq:SINR-value}.

\vspace{3mm}

\textbf{Proof of Theorem \ref{theorem:average-SE}.} We compute a lower bound on $\mathbb{E} \{ \mathrm{SE}_{0k} \} $, for $\mathrm{SE}_{0k}$ given in Theorem \ref{theorem:SINR-MRC}, by using Jensen's inequality as
\begin{equation}
\mathbb{E} \{ \mathrm{SE}_{0k} \} \geq  A \Big( 1 - \frac{B}{S} \Big) \log_2 \! \left( 1 + \mathbb{E} \{ \mathrm{SINR}_{0k}^{-1}\}  \right).
\end{equation}
It remains to compute all expectations that appear in $\mathbb{E} \{ \mathrm{SINR}_{0k}^{-1}\}$.
When applying the power-control from \eqref{eq:power-control}, we obtain an expression only containing the expectations
\\
\begin{equation}
\mathbb{E} \Bigg\{  \sum\limits_{j \in \Psi} \sum\limits_{i=1}^{K}  \frac{d_{jji}^{\gamma \alpha}}{d_{0ji}^{\gamma\alpha}}   \Bigg\} = \begin{cases}
\frac{2K}{\alpha-2}, & \gamma=1, \\
\frac{K}{\alpha-1},& \gamma = 2,
\end{cases}
\end{equation}
\begin{equation} \label{eq:second-moment}
\mathbb{E} \Bigg\{ \Bigg( \sum\limits_{j \in \Psi} \sum\limits_{i=1}^{K}  \frac{d_{jji}^{ \alpha}}{d_{0ji}^{\alpha}}   \Bigg)^2 \Bigg\} \leq   \frac{4K^2}{(\alpha-2)^2} + \frac{K^2}{\alpha-1} ,
\end{equation}
which were computed in \cite[Proof of Prop.~1]{Bjornson2016d}. The inequality in \eqref{eq:second-moment} further establishes \eqref{eq:SINR-value-stochastic} as a lower bound.

\bibliographystyle{IEEEbib}
\bibliography{IEEEabrv,refs}

\end{document}